%% file: main.tex
\documentclass[10pt,twocolumn,letterpaper]{article}

\usepackage{cvpr}              
\usepackage{times}
\usepackage{epsfig}
\usepackage{graphicx}
\usepackage{amsmath}
\usepackage{amssymb}
\usepackage{booktabs}
\usepackage{rotating}
\usepackage{subcaption}
\usepackage{multirow}
\usepackage{cite}
\input{preamble}

\definecolor{cvprblue}{rgb}{0.21,0.49,0.74}
\newcommand{\myccone}{\cellcolor[HTML]{f2f2f2}}
\usepackage[pagebackref,breaklinks,colorlinks,citecolor=cvprblue]{hyperref}

\def\paperID{12269} 
\def\confName{CVPR}
\def\confYear{2024}

\title{AV-RIR: Audio-Visual Room Impulse Response Estimation}

\author{Anton Ratnarajah \quad Sreyan Ghosh \quad Sonal Kumar \quad Purva Chiniya \quad Dinesh Manocha\\
University of Maryland, College Park\\
{\tt \{jeran,sreyang,sonalkum,pchiniya,dmanocha\}@umd.edu}}

\begin{document}

\maketitle
\input{sec/0_abstract}    
\input{sec/1_intro}
\input{sec/2_realted-works}

\input{sec/3_approach}
\input{sec/4_experiment}
\input{sec/5_conclusion}
\input{sec/6_supplementary}
{
    \small
    \bibliographystyle{ieeenat_fullname}
    \bibliography{main}
}


\end{document}

%% file: preamble.tex
\usepackage[dvipsnames,table]{xcolor}


%% file: sec/0_abstract.tex
\begin{abstract}



Accurate estimation of Room Impulse Response (RIR), which captures an environment's acoustic properties, is important for speech processing and AR/VR applications. We propose AV-RIR, a novel multi-modal multi-task learning approach to accurately estimate the RIR from a given reverberant speech signal and the visual cues of its corresponding environment. AV-RIR builds on a novel neural codec-based architecture that effectively captures environment geometry and materials properties and solves speech dereverberation as an auxiliary task by using multi-task learning.  We also propose Geo-Mat features that augment material information into visual cues and CRIP that improves late reverberation components in the estimated RIR via image-to-RIR retrieval by 86\%. Empirical results show that AV-RIR quantitatively outperforms previous audio-only and visual-only approaches by achieving 36\% - 63\% improvement across various acoustic metrics in RIR estimation. Additionally, it also achieves higher preference scores in human evaluation. As an auxiliary benefit, dereverbed speech from AV-RIR shows competitive performance with the state-of-the-art in various spoken language processing tasks and outperforms reverberation time error score in the real-world AVSpeech dataset. Qualitative examples of both synthesized reverberant speech and enhanced speech are available online\footnotemark{}. 

\footnotetext{\url{https://anton-jeran.github.io/AVRIR/}}





\end{abstract}


%% file: sec/1_intro.tex
\section{Introduction}
\label{sec:intro}

Reverberation, caused by sound reflecting off surrounding surfaces, transforms how a listener perceives the sound once it is released from a sound source. The transformation is influenced by specific properties of the surrounding area, like spatial geometry, the composition and material properties of surfaces and objects within the environment, and the positioning of various sound sources in proximity. For example, someone speaking or playing music in a large auditorium is perceptually significantly different from someone speaking in a small classroom~\cite{auditorium,small_room}. The environmental effect that any sound goes through because of the transformation can be quantitatively described by the room impulse response (RIR). RIR is a fundamental concept that characterizes how an acoustic space affects sound, essentially representing the transfer function between a sound source and a receiver, encapsulating all the direct and reflective paths that sound can travel within any indoor or outdoor environment. 

\begin{figure}[t] 
\centering
\includegraphics[width=\columnwidth]{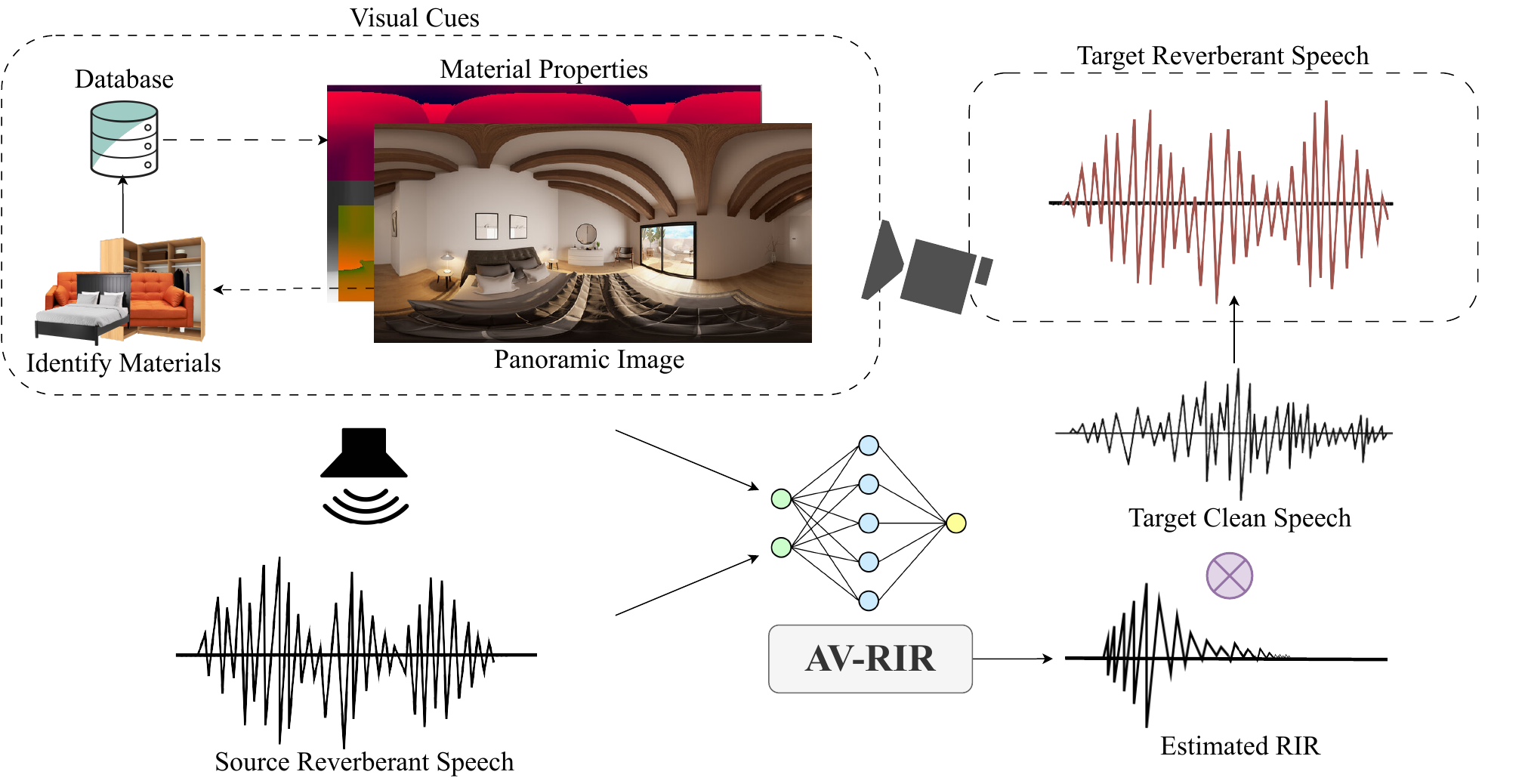}
\caption{\small Overview of \textbf{AV-RIR}: Given a source reverberant speech in any environment, AV-RIR estimates the RIR from the reverberant speech using additional visual cues. The estimated RIR can be used to transform any target clean speech as if it is spoken in that environment.} 
\label{feaute_map}
\vspace{-0.3cm}
\end{figure}

RIR estimation, defined as estimating the RIR component from a given reveberant speech signal (see Eq. \ref{eqtn:reverberant_speech}), finds its major application in augmented reality (AR) and virtual reality (VR) \cite{projectacoustic,steam-audio,oculus-spatializer,liu2022sound}. Usually, when sound effects do not align acoustically with the visual scene, it can disrupt the audio-visual human perception. In AR and VR settings, discrepancies between the acoustics of the real environment and the virtually simulated space lead to cognitive dissonance. This phenomenon, known as the ``room divergence effect'', can significantly detract from the user experience \cite{raghuvanshi2022interactive,scene-aware}. RIR estimation from real-world speech can help overcome these problems.
Prior work on RIR estimation mainly deals with recorded 
 audio signals and does not take into account visual cues. Directly estimating the RIR from source reverberant speech has been extensively studied using traditional signal processing methods~\cite{esti_mic1,esti_mic2,esti_gaus1,esti_gaus2}. However, these approaches may not work well in some real-world applications, mainly because they are based on the assumption that the source is a modulated Gaussian pulse, not actual speech, ~\cite{esti_gaus1,esti_gaus2} or they require pre-knowledge of the specific attributes about the speaker or the microphone used for recording~\cite{esti_mic1,esti_mic2}. Recently, neural learning-based RIR estimation techniques have been proposed to estimate RIR from reverberant speech~\cite{fins,s2ir}. These techniques are capable of estimating early components (i.e., the direct response and early reflections of RIR) and are not very effective in estimating late components (i.e., the late reverberation of RIR) because the early components of the RIR have impulsive sparse components, while the late components have a noise-like structure with significantly lower magnitude compared to early components. Therefore audio-only approach approximates the late components using a sum of decaying filtered noise signal~\cite{fins,moorer}. 



\vspace{0.5mm}


\looseness=-1
{\noindent \textbf{Main Contributions.}} We propose AV-RIR, a novel multi-modal multi-task learning approach for RIR estimation. AV-RIR employs a novel neural codec-based multi-modal architecture that takes as input a reverberant speech uttered in a source environment, the panoramic image of the environment, and a novel Geo-Mat feature that incorporates information about room geometry and the materials of surfaces and objects. The multi-modal architecture consists of carefully designed encoders, decoders, and a Residual Vector Quantizer that learns rich task-specific (RIR estimation and speech deverberation) features while discarding the noise in training data~\cite{lip_movement}. Additionally, AV-RIR incorporates a dual-branch structure for multi-task learning, where we solve an auxiliary speech dereverberation task alongside the primary RIR estimation task. This approach effectively redefines the ultimate learning objective as decomposing reverberated speech into its constituent anechoic speech and RIR components. Furthermore, we propose Contrastive RIR-Image Pre-training (CRIP) to improve late reverberation of the estimated RIR during inference time using image-to-RIR retrieval. To summarize, our main contributions are as follows:

\begin{enumerate}
    \item We propose AV-RIR, a novel multi-modal multitask learning approach for RIR estimation. 
    
    \item AV-RIR employs a neural codec-based multi-modal architecture that takes as input audio, visual cues and a novel Geo-Mat feature. We also propose CRIP to improve late reverberation effects using retrieval.
    

    \item During training, AV-RIR solves an auxiliary speech dereverberation task for learning RIR estimation. Through this, AV-RIR essentially learns to separate anechoic speech and RIR.
    
    \item We perform extensive experiments to prove the effectiveness of AV-RIR. AV-RIR outperforms prior works by significant margins both quantitatively and qualitatively. We achieve 36\% - 63\% on RIR estimation on the SoundSpaces dataset~\cite{soundspaces1}, and 56\% - 79\% people find that AV-RIR is closer to the ground-truth in the visual acoustic matching task over our baselines. Additionally, the dereverbed speech predicted by AV-RIR improves performance across various spoken language processing (SLP) tasks. We also perform extensive ablation experiments to demonstrate the critical role of each modules within the AV-RIR framework.
\end{enumerate}

%% file: sec/2_realted-works.tex
\begin{figure*}[t] 
\centering
\includegraphics[width=0.95\linewidth]{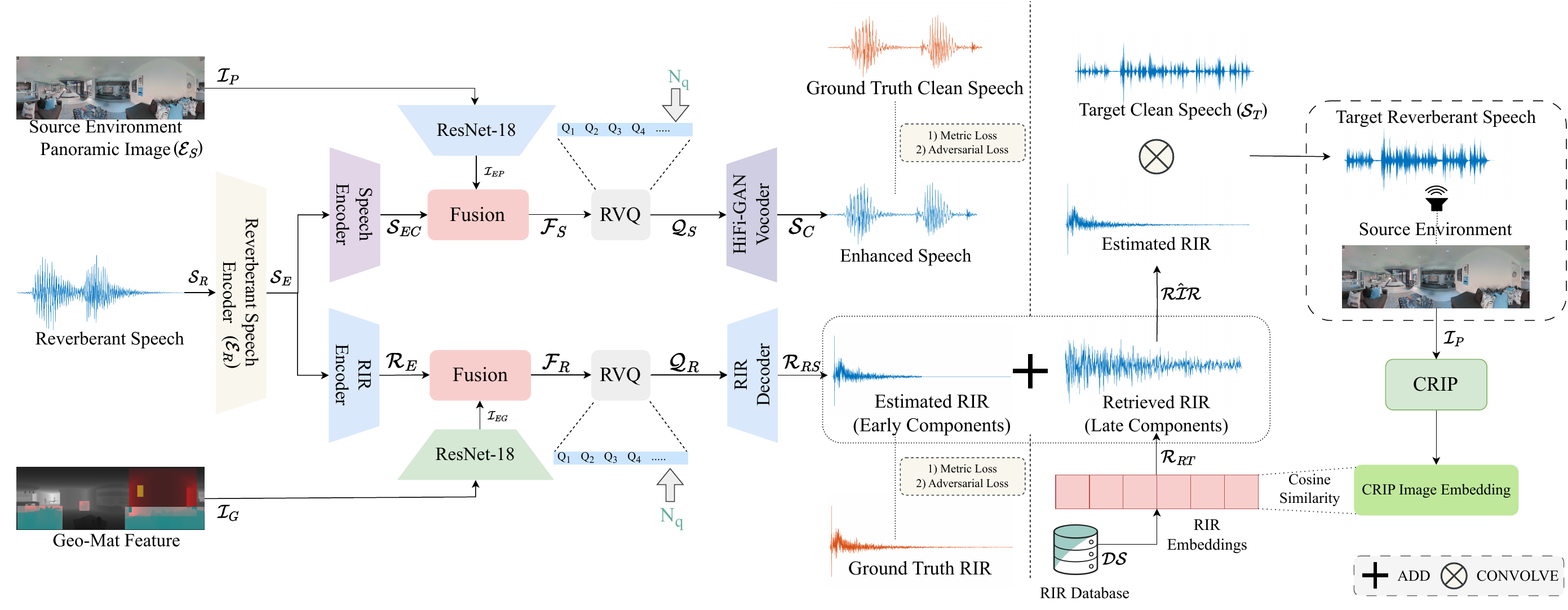}
\caption{\small Overview of our AV-RIR learning method: Given the input reverberant speech $\mathcal{S}_R$ from any source environment $\mathcal{E}_{S}$, the primary task of AV-RIR is to estimate the room impulse response $\mathcal{RIR}$ by separating it from the clean speech $\mathcal{S}_C$ (see Eq.~\ref{eqtn:reverberant_speech}). The input $\mathcal{S}_R$ is first encoded using a Reverberant Speech Encoder $\mathcal{E}_{{R}}$. The latent output $\mathcal{S}_E$ is then passed to two different encoders in two different branches. While one of these branches solves the RIR estimation task, the other solves the speech dereverberation task by estimating $\mathcal{S}_C$. Outputs from both the Speech Dereverberation Encoder $\mathcal{S}_{EC}$ and RIR Encoder $\mathcal{R}_{E}$ are fused with ResNet-18 encodings from the panoramic image $\mathcal{I}_{P}$ and Geo-Mat feature $\mathcal{I}_{G}$ respectively. The output latent multi-modal encodings $\mathcal{I}_{EP}$ and $\mathcal{I}_{EG}$ are then passed to a trainable Residual Vector Quantization module (RVQ), which quantizes $\mathcal{F}_{S}$ to latent codes $\mathcal{Q}_{S}$, and $\mathcal{F}_{R}$ to latent codes $\mathcal{Q}_{R}$. Finally, the HiFi-GAN vocoder decodes the enhanced speech $\mathcal{S}_C$ from $\mathcal{Q}_{S}$ and the RIR decoder decodes estimated early components of RIR $\mathcal{R_{RS}}$ from $\mathcal{Q}_{R}$ which are used to calculate losses for training. At inference time, our CRIP retrieves an RIR from a database $\mathcal{DS}$ and is used to improve late reverberation in the estimated RIR. Finally, post addition, the final estimated RIR is convolved with any $\mathcal{S}_C$ to make it sound like it was uttered in $\mathcal{E}_{S}$.}

\label{fig:full_archi}
\vspace{-0.3cm}
\end{figure*}
\section{Background and Related Work}
\label{sec:related}

{\noindent \textbf{Room Impulse Response.}}
The reverberation effects in a recorded speech can be characterized by a transfer function known as room impulse response (RIR)~\cite{irgan}. We can breakdown the speech content ($\mathcal{S}_C$) and the RIR ($\mathcal{RIR}$) corresponding to the given reverberant speech ($\mathcal{S}_R$) using a convolution operation ($\circledast$) as follows:
{
\belowdisplayskip 0.8\belowdisplayshortskip
\begin{equation}\label{eqtn:reverberant_speech} 
\begin{aligned}[b]
     \operatorname{\mathcal{S}_R = \mathcal{S}_C \ \circledast \ \mathcal{RIR},}
\end{aligned}
\end{equation}
} 
\noindent where RIR represents the intensity and time of arrival of direct sound, early reflections, and late reverberation. The RIR can either be measured in a controlled environment~\cite{real_rir4,real_rir5,real_rir3,chen2024RAF} or simulated using physics-based simulators~\cite{precompute1,dynamic2}. Measuring RIR requires sophisticated hardware and human labor. At the same time, acoustic simulators require a 3D mesh representation of the environment~\cite{GWA} and complete knowledge of room materials~\cite{acoustic_classification}. Thus, these simulators cannot capture all the acoustic effects of an RIR at an interactive rate~\cite{GWA}. Prior works also show that accurately generating RIRs without these cues is infeasible (e.g., only from RGB images~\cite{image2reverb,fewshot-rir}). Alternatively, estimating RIR from reverberant speech has seen better success~\cite{fins}. Reverberant speech can be easily recorded using household devices (e.g., mobile phone, Amazon Echo, etc.). Prior work explores audio-only techniques~\cite{s2ir,fins,yet_rir}, and to the best of our knowledge, AV-RIR is the first audio-visual method to solve this task.
\vspace{0.5mm}

{\noindent \textbf{Room Acoustic Estimation and Matching.}} Interactive applications (i.e., AR / VR, computer games, etc.) demand accurate RIRs to generate realistic sound effects. Many physics-based solutions have been proposed to generate RIRs for synthetic scenes~\cite{precompute1,precompute2,precompute3,dynamic1,dynamic2}. Alternatively, several machine-learning algorithms are being proposed to estimate RIRs for the given environment~\cite{NAF1,fewshot-rir,mesh2ir,fast-rir,be-everywhere,NAF2,listen2scene}. Pioneering learning-based work on generating RIR from a single RGB image of physical environments includes Image2Reverb~\cite{image2reverb}, which is based on a conditional GAN-based architecture. Generating RIR from a single RGB image might not be the most effective as it does not have enough information, like 3D geometry, information about the material properties of objects in the environment, speaker position, etc. The existing physics-based and machine-learning solutions to generate accurate RIRs for dynamically moving listeners and sources use a 3D geometric representation of the environment, the locations of the speaker and listener, and a few measured RIRs from the same real environment. In many real-world scenarios, we do not have the 3D representation of the environment and measured RIR from the same environment. 


Recent works  demonstrate the feasibility of estimating early RIR components from reverberant speech~\cite{s2ir,fins}, and predicting late reverberation from the image of the environment~\cite{image2reverb}. Furthermore, recent advances involve neural algorithms for converting reverberant sounds between different environments~\cite{visual_acoustic1,visual_acoustic2,AV-NeRF,novel-acoustic}, diverging from traditional methods reliant on pre-computed RIRs as described in Eq.~\ref{eqtn:reverberant_speech}. However, these end-to-end approaches often incur substantial latency due to deep model inference, making them less viable for real-time mobile applications.



\vspace{0.5mm}

{\noindent \textbf{Speech Dereverberation.}} The human auditory cortex's ability to adaptively filter out reverberation in various acoustic environments is well-documented~\cite{auditory_cortex1}. Inspired by this, researchers have developed speech enhancement systems capable of transforming reverberant speech to anechoic speech~\cite{compute_reverb1,drr_book,compute_reverb3}. Initially focused on multi-microphone inputs~\cite{multi1,multi2,multi3}, recent deep learning techniques have shown promise with single-channel inputs~\cite{single1,skip-convgan,tcnn,cross-domain}. While visual cues have been explored for speech enhancement, most studies have concentrated on near-field ASR using visible lip movements of the speaker~\cite{lip_movement,Visual_Speech1,Visual_Speech2}. Recent works use panoramic room images for speech dereverberation~\cite{AV_changan,AV_sanjoy}.

%% file: sec/3_approach.tex
\section{Methodology}
\label{sec:method}

\subsection{Overview: AV-RIR}
\label{sec:overview}
\begin{figure}[t] 
\centering
\includegraphics[width=\columnwidth]{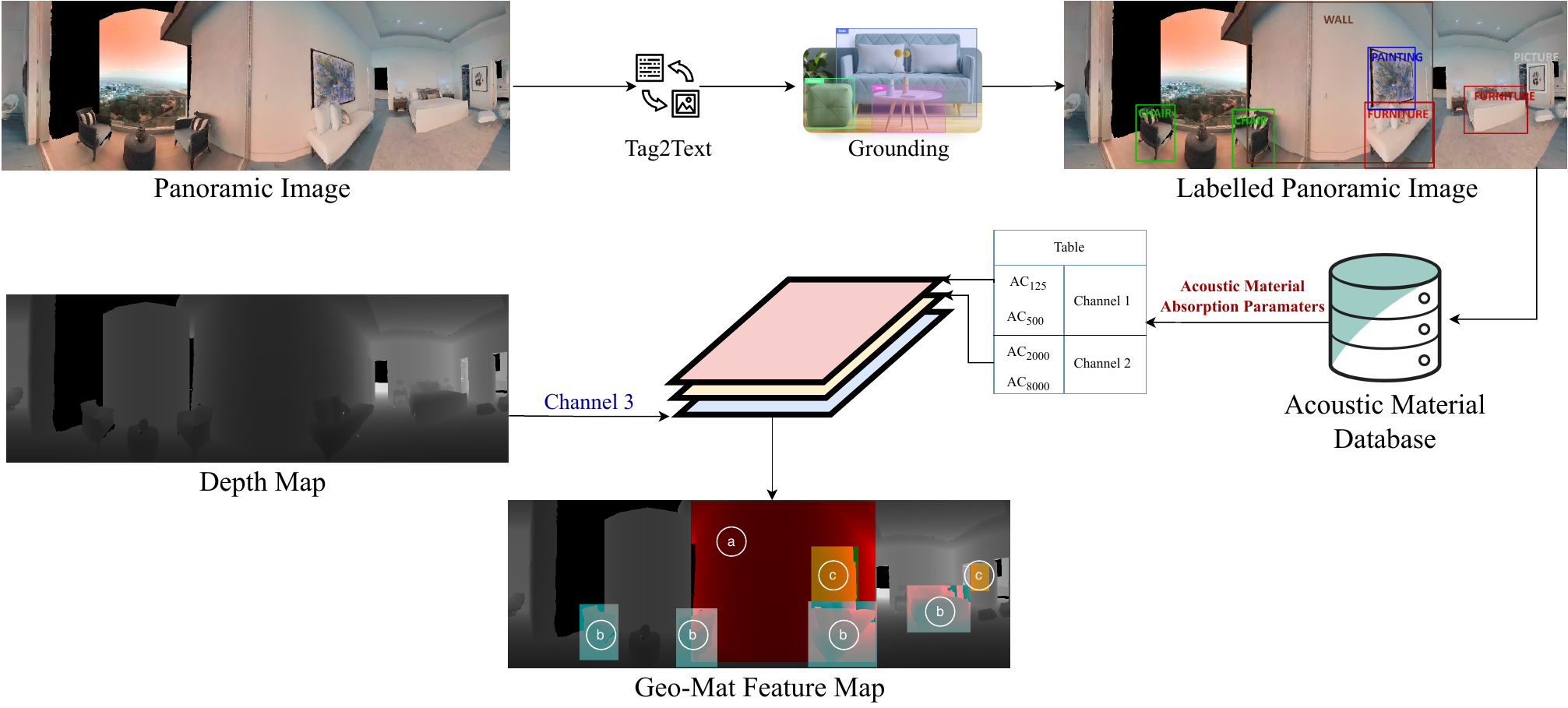}
\caption{\small The computation pipeline of \textbf{Geo-Mat feature map}. The first two channels of the Geo-Mat feature ($\mathcal{I}_{G}$) comprise the absorption coefficients ($\mathcal{AC}$) of each acoustic material. The third channel comprises the depth map. We illustrate objects in the environment having similar $\mathcal{AC}$ with similar colors:  chairs and furniture with similar materials are represented in light blue, painting, and wall pictures with similar materials are represented in yellow, and the rest in grey. More details on the method to obtain $\mathcal{AC}$ is described in Section~\ref{subsec:Geo-Mat}.} 
\label{Geo-Mat}
\vspace{-0.3cm}
\end{figure}

\begin{figure}[t] 
\centering
\includegraphics[width=0.9\columnwidth]{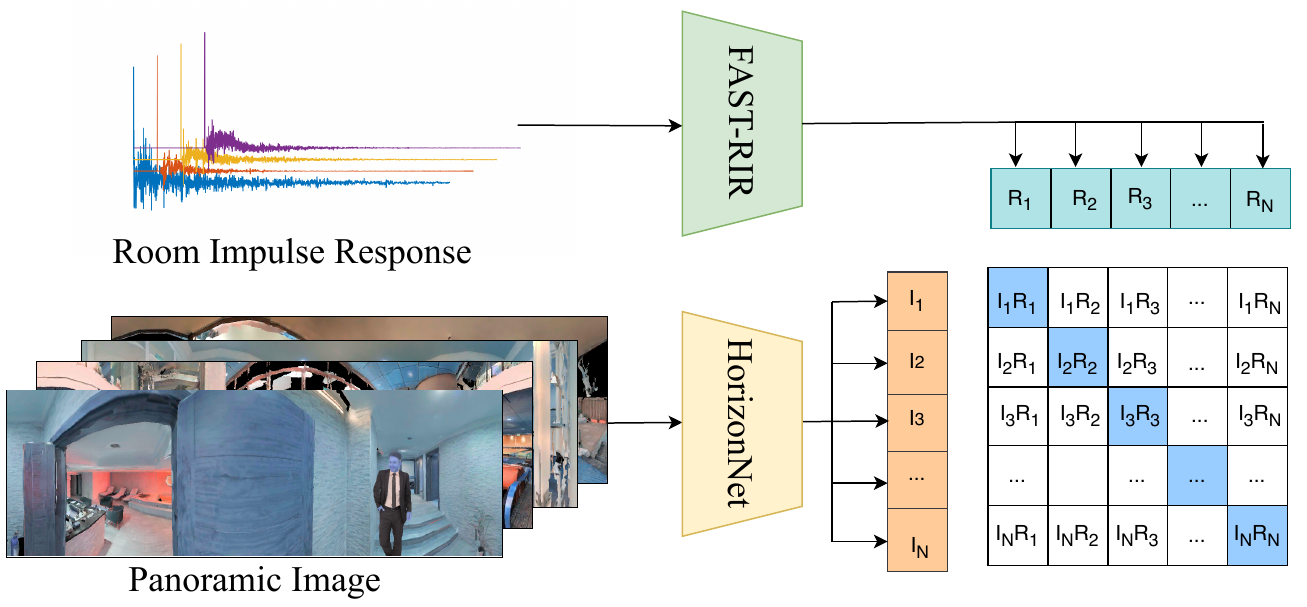}
\caption{\small Illustration of \textbf{CRIP} training. Like CLIP~\cite{clip}, we propose two networks, one to encode a panoramic image and the other to encode the RIR to learn a joint embedding space between both. We use our CRIP-based image-to-RIR retrieval during inference to improve late reverberation in the estimated RIR from AV-RIR.} 
\label{fig:CRIP}
\vspace{-0.5cm}
\end{figure}

Fig.~\ref{fig:full_archi} gives an overview of our approach. Given a reverberant speech $\mathcal{S}_R$, the task of AV-RIR is to learn to estimate the RIR from $\mathcal{S}_R$. To achieve this, we propose a novel multi-modal neural architecture and solve two parallel tasks for learning accurate RIR estimation. As input, together with the $\mathcal{S}_R$, AV-RIR also receives the RGB panoramic image $\mathcal{I}_P$ of the source environment and our proposed Geo-Mat feature map $\mathcal{I}_G$. The construction of $\mathcal{I}_G$ is illustrated in Fig.~\ref{Geo-Mat}. The $\mathcal{S}_R$ is first passed through a reverberant speech encoder, after which AV-RIR breaks into two branches that solve two different tasks. The bottom branch, which also receives ResNet-18 encoded features of $\mathcal{I}_{G}$, solves our primary RIR estimation task. The other branch, which receives the encoded features of $\mathcal{I}_P$, solves an auxiliary speech dereverberation task to predict enhanced speech $\mathcal{S}_C$. After the multi-modal feature fusion step, both branches employ a Residual Vector Quantizer module. During inference, to synthesize any target speech as if spoken in the source environment, we convolve the estimated RIR with the target clean speech. Additionally, we propose CRIP (Fig.~\ref{fig:CRIP}) to retrieve an RIR from a datastore $\mathcal{DS}$, conditioned on the $\mathcal{I}_P$ of the source environment, to improve late components. Next, we will describe each module in detail.




\subsection{AV-RIR Architecture}
\label{subsec:arc}

{\noindent \textbf{Reverberant Speech Encoder  ($\mathcal{E}_{{R}}$).}} Our $\mathcal{E}_{{R}}$ consists of a simple CNN-based architecture with a single 1-D CNN layer and a single input and output channel. As speech dereverberation and RIR estimation are similar learning problems and based on convolution operations (the latter learns deconvolution, and the former learns the inverse), the latent output $\mathcal{S}_{E}$ from the encoder serves as efficient representations for both tasks AV-RIR solves.
\vspace{0.5mm}

{\noindent \textbf{Room Impulse Response Encoder ($\mathcal{E}_{{IR}}$).}} $\mathcal{E}_{{IR}}$ is adapted and modified from the S2IR-GAN encoder~\cite{s2ir}. Our three-layer $\mathcal{E}_{{IR}}$ has 256, 512, and 1024 output channels, 14401, 41, and 41 kernel lengths, and 225, 2, and 2 strides, respectively. The large kernel length in the first layer encodes the RIR features efficiently. We significantly reduce the input dimension by a factor of 900. We process reverberant speech $\mathcal{S}_{R}$ segments of $\mathbb{R}^{1 \times 14400}$ samples. Therefore, every reverberant speech sample is encoded into $\mathbb{R}^{1024 \times 16}$ RIR temporal features.
\vspace{0.5mm}

{\noindent \textbf{Vision Encoders ($\mathcal{E}_{{P}}$, $\mathcal{E}_{{GM}}$).}} Prior work on audio-visual speech dereverberation and localization has shown that ResNet-18~\cite{resnet} is capable of extracting strong cues from image $\mathcal{I}_{P}$ and depth maps~\cite{AV_changan,fewshot-rir,novel-acoustic}. Therefore, we employ two separate ResNet-18-based feature encoders $\mathcal{E}_{{P}}$ and $\mathcal{E}_{{GM}}$ to encode the $\mathcal{I}_P$ and the Geo-Mat feature $\mathcal{I}_G$, respectively, and reshape the features to $\mathbb{R}^{1024\times4}$. 

\vspace{0.5mm}

{\noindent \textbf{Multi-modal Fusion Modules ($\mathcal{M}$).}}
Similar to previous neural audio codec architectures~\cite{lip_movement}, we fuse the visual features with the audio stream along the temporal axis. The combined audio-visual encoded representation is projected into the designed multi-dimensional space~\cite{audiodec} and passed to the next stage to quantize into codes.
\vspace{0.5mm}

{\noindent \textbf{Residual Vector Quantizer (RVQ).}} RVQ are used in neural audio codecs to compress the encoder output into a discrete set of code vectors to transmit the data at a fixed target bitrate $\mathcal{R}$ (bits/second). For AV-RIR, we modify the RVQ proposed in SoundStream~\cite{SoundStream}. SoundStream proposes a VQ with trainable codebooks that are trained together with the model end-to-end. SoundStream~\cite{SoundStream} adapts the VQ proposed in~\cite{VQ1,VQ2} and improves the codebook with the multi-stage VQ~\cite{RVQ}. Our primary modification involves relaxing the constraints (i.e., the target bitrate) to improve the speech dereverberation performance. SoundStream is designed for real-time transmissions and streaming compressed audio at 3-18 Kbps. For our task, we relaxed the compression to $\approx$59 Kbps. We observed that relaxing beyond 59 Kbps did not significantly improve the performance. Audio codecs have shown increased performance by increasing the birates in subjective tests~\cite{encodec}. Our RVQ cascades $N_{q} = 64$ layers of VQ and uses a large codebook size $\mathcal{N} = 8192$. Having a larger $\mathcal{N} / \mathcal{N}_{q}$ ratio has been shown to achieve higher coding efficiency~\cite{SoundStream}.

\vspace{0.5mm}

{\noindent \textbf{Decoders ($\mathcal{D}_{IR}$, $\mathcal{D}_{S}$)}.} As the two branches of AV-RIR are responsible for decoding separate outputs from the compressed codes (i.e., enhanced speech and RIR), we use two different decoders for this task. We use a HiFi-GAN vocoder for the speech dereverberation branch to decode enhanced speech from the compressed code. HiFi-GAN~\cite{hi-fi_vocoder} has shown impressive performance in generating high-fidelity speech, especially with audio codecs~\cite{audiodec}. For the RIR estimation branch, we use a modified SoundStream decoder~\cite{SoundStream}. We modify the decoder to have 6 transposed convolutional (Conv) blocks with output channels of (256, 128, 64, 32, 32, 16) and strides of (5, 5, 2, 2, 1, 1). The output from the last transposed Conv block is passed to a final 1D Conv layer with kernel size 1 and stride 1 to project the code to the waveform domain.
\vspace{0.5mm}

\subsection{Geo-Mat Features}
\label{subsec:Geo-Mat}
The Geo-Mat feature ($\mathcal{I}_{G}$) represents the geometry and sound absorption properties of materials in the environment. Physics-based RIR simulators take the 3D geometry of the environment and the material absorption coefficients ($\mathcal{AC}$) of each material present in the environment as input to accurately generate RIR for that environment~\cite{wave1,interactive1,soundspaces2}. On the other hand, Changhan \textit{et al.}~\cite{AV_changan} show that leveraging depth maps improves the performance of audio-visual dereverberation. Inspired by these techniques, we propose Geo-Mat $\mathcal{I}_{G}$ to improve AV-RIR's understanding of geometry and material information of the environment. Our proposed $\mathcal{I}_{G}$ is a 3-channel feature map constructed before training our AV-RIR using $\mathcal{AC}$ as the first two channels and the depth map as the last channel. We represent $\mathcal{I}_{G}$ with 3 channels to encode $\mathcal{I}_{G}$ using commonly used image encoder~\cite{resnet}.


\vspace{0.5mm}

{\noindent \textbf{Obtaining material absorption coefficients ($\mathcal{AC}$).}} To obtain absorption coefficients $\mathcal{AC}$ of all materials in the environment from image $\mathcal{I}_{P}$, we employ a language-guided pipeline with SOTA pre-trained models. We first use Tag-2-Text~\cite{huang2023tag2text}, a SOTA object tagging model that identifies all objects in the image. This is followed by Grounding DINO~\cite{sag1,sag2}, which provides bonding box locations of each object identified by Tag-2-Text. Next, we use a large-scale room acoustic database with measured frequency-dependent $\mathcal{AC}$ to match the $\mathcal{AC}$ for each detected object~\cite{kling_2018}. For the matching operation, we adhere to a simple semantic matching technique to match the material names in the database to the ones detected using Tag-2-Text using embeddings from sentence transformer~\cite{sentence_bert}. Precisely, we calculate an embedding $e_{\mathcal{I}_P} \in \mathbb{R}^{768}$ for every object detected by Tag-2-Text and an embedding $e_{\mathcal{M}} \in \mathbb{R}^{768}$ for every object in the database. Then we take the coefficient of the material in the database with the highest cosine similarity to $e_{\mathcal{I}_P}$. The acoustic $\mathcal{AC}$ are frequency-dependent~\cite{GWA} and, therefore, we use the sub-band $\mathcal{AC}$ at 125 Hz, 500 Hz, 2000 Hz, and 8000 Hz to create the Geo-Mat feature $\mathcal{I}_{G}$.
\vspace{0.5mm}

{\noindent \textbf{Feature Map Construction.}} After obtaining the absorption coefficients ($\mathcal{AC}$) for each material in the panoramic image ($\mathcal{I}_{P}$), we finally construct our 3-channel Geo-Mat feature $\mathcal{I}_{G}$ (Eq.~\ref{Geo-Mat1}). \textbf{Channel 1:} The material $\mathcal{AC}$ at low frequencies (i.e., 125 Hz and 500 Hz)  \textbf{Channel 2:} The $\mathcal{AC}$ of the materials at high frequencies (i.e., 2000 Hz and 8000 Hz). \textbf{Channel 3:} The last channel of the $\mathcal{I}_{G}$ represents the monocular depth map ($\mathcal{I}_{D}$) of the $\mathcal{I}_{P}$. Most datasets used in our experiments provide depth maps; however, for datasets that do not, we use the system provided by Godard \textit{et al.}~\cite{depth_estimate} to compute the depth map from a single image. We notice that the order of the channels does not matter.
{
\belowdisplayskip 0.8\belowdisplayshortskip
\begin{equation}\label{Geo-Mat1} 
\small
\begin{aligned}[b]
   & \operatorname{\mathcal{I}_{G}}[:,:,0] = \mathcal{AC}_{125} + \mathcal{AC}_{500}*16 \\
    &\operatorname{\mathcal{I}_{G}}[:,:,1] = \mathcal{AC}_{2000} + \mathcal{AC}_{8000}*16 \\
    &\operatorname{\mathcal{I}_{G}}[:,:,2] = \mathcal{I}_{D}
\end{aligned}
\end{equation}
}




\subsection{Training AV-RIR}
\label{{sec:training_objective}}

$\mathcal{D}_{{QIR}}$ and $\mathcal{D}_{{QS}}$ are RIR and clean speech quantizers followed by a decoder, respectively, and $\mathcal{CRIP}$ represents CRIP. We estimate clean speech ($\hat{\mathcal{S}_{{C}}}$), RIR ($\hat{\mathcal{RIR}}$), and reverberant speech ($\hat{\mathcal{S}_{{R}}}$) as shown below.
{\small
 \belowdisplayskip 0.4\belowdisplayshortskip
\begin{equation}\label{eq:estimations_clean}
\begin{aligned}[b]
   &\operatorname{\hat{\mathcal{S}_{C}} = \mathcal{D}_{QS}(\mathcal{E}_{P}(\mathcal{I}_{P}), \  \mathcal{E}_{S}(\mathcal{S}_{R})).}\\
    &\operatorname{\hat{\mathcal{RIR}} = \mathcal{D}_{QIR}(\mathcal{E}_{GM}(I_{GM}), \ \mathcal{E}_{IR}(\mathcal{S}_{R})) \ + \mathcal{CRIP}(I_{P}).}\\
    &\operatorname{\hat{\mathcal{S}_{R}} =  \hat{\mathcal{S}_{C}} \circledast \hat{\mathcal{RIR}}.}
\end{aligned}
\end{equation}
}


{\noindent \textbf{RIR Estimation Loss.}}
We calculate the time-domain mean squared error (MSE) for the estimated RIR as follows.
{
\small
 \belowdisplayskip 0.4\belowdisplayshortskip
\begin{equation}\label{rir_loss}
\begin{aligned}[b]
     \operatorname{\mathcal{L}_{MSE} = \mathbb{E}[||\mathcal{RIR} \ - \ \hat{\mathcal{RIR}}||_{2}] .}
\end{aligned}
\end{equation}
}

To train our RIR estimation RVQ codebook, we use the exponential moving average loss proposed in~\cite{VQ1} as our vector quantizer (VQ) loss $\mathcal{L}_{{VQ1}}$.
 \vspace{0.5mm}
 
{\noindent \textbf{Speech Dereverberation Loss.}} For solving the speech dereverberation task, we optimize three losses: 
 \vspace{0.5mm}
 
{\noindent \textbf{(1) Mel-Spectrogram (Mel) loss ($\mathbf{\mathcal{L}_{{Mel}}}$).}} The mel-spectrogram loss helps improve the perceptual quality of the predicted enhanced speech~\cite{post-filter}. The 1D waveform $(\hat{\mathcal{S}_{{R}}})$ output from the HiFi-GAN vocoder is first converted to the Mel-spectrogram, transforming it from the time domain to the frequency domain representation. A drawback of the Mel loss is its fixed resolution. The window length determines whether it has a good frequency or time resolution (i.e., a wide window has good frequency resolution and poor time resolution). Therefore, we calculate $\mathcal{L}_{{Mel}}$, over a range of window lengths $W_L$ = \{64, 128, 256, 512, 1024, 2048, 4096\}. Formally, $\mathcal{L}_{{Mel}}$ can be defined as:
{
\small
 \belowdisplayskip 0.4\belowdisplayshortskip
\begin{equation}\label{mel_loss}
\begin{aligned}[b]
    \operatorname{\mathcal{L}_{MEL} = \mathbb{E}[||MEL(\mathcal{S}_{R}) - MEL(\hat{\mathcal{S}_{R}}) )||_{1} +} \\
    \operatorname{||MEL(\mathcal{S}_{C}) - MEL(\hat{\mathcal{S}_{C}}) )||_{1}],}
\end{aligned}
\end{equation}
}

where $\operatorname{MEL()}$ is the operation that converts time-domain speech into its mel-spectrogram representation. 
\vspace{0.5mm}

{\noindent \textbf{(2) Short-Time Fourier Transform (STFT) loss ($\mathbf{\mathcal{L}_{STFT}}$}).} The STFT loss helps in the high-fidelity reconstruction of the predicted enhanced speech~\cite{AV_changan}. The 1D waveform is first converted to the frequency domain by applying the $\operatorname{STFT()}$ operation. The STFT of a waveform can be represented as a complex spectrogram where $\mathcal{M}_{{S}}$ represents the magnitude of the STFT and $\mathcal{P}_{{S}}$ is the phase of the STFT. We map the phase angle of the STFT  to the rectangular coordinate on the unit circle to avoid phase wraparound issues~\cite{AV_changan}. Our $\mathcal{L}_{STFT}$ is the sum of magnitude loss $\mathcal{L}_{{MAG}}$ and phase loss $\mathcal{L}_{{PH}}$. Similar to $\mathcal{L}_{{Mel}}$, we calculate $\mathcal{L}_{STFT}$ over a range of window lengths in a similar setting.
{
\small
 \belowdisplayskip 0.4\belowdisplayshortskip
\begin{equation}\label{stft_mag_loss}
\begin{aligned}[b]
    & \operatorname{\mathcal{L}_{MAG} = \mathbb{E}[||M_{S}(\mathcal{S}_{R}) - M_{S}(\hat{\mathcal{S}_{R}}) )||_{2} + ||M_{S}(\mathcal{S}_{C}) - M_{S}(\hat{\mathcal{S}_{C}}) )||_{2} ].} \\
     & \operatorname{\mathcal{L}_{P} (x,\hat{x}) = \mathbb{E}[||\sin{(P_{S}(x))}- \sin{(P_{S}(\hat{x})))}||_{2}} \\
      & \operatorname{\quad \quad \quad \quad \quad+ ||\cos{(P_{S}(x))}- \cos{(P_{S}(\hat{x}))}||_{2} ].} \\
     &\operatorname{ \mathcal{L}_{PH} = \mathcal{L}_{P} ((\mathcal{S}_{R},\hat{(\mathcal{S}_{R}}) + \mathcal{L}_{P} (\mathcal{S}_{C},\hat{\mathcal{S}_{C}}).} \\
     &\operatorname{ \mathcal{L}_{STFT} = \mathcal{L}_{MAG} + \mathcal{L}_{PH}}
\end{aligned}
\end{equation}
}

     

Our total metric loss $\mathcal{L}_{METRIC}$ (including RIR estimation and speech derverberation) is described in Eq.~\ref{metric_loss}. 
{
\small
 \belowdisplayskip 0.4\belowdisplayshortskip
\begin{equation}\label{metric_loss}
\begin{aligned}[b]
     \operatorname{\mathcal{L}_{METRIC} = \mathcal{L}_{MEL} + \lambda_{1} \mathcal{L}_{STFT} + \lambda_{2} \mathcal{L}_{MSE},}
\end{aligned}
\end{equation}
}
where $\lambda_{1}$and $\lambda_{2}$ are the weights. 
\vspace{0.5mm}

{\noindent \textbf{(3) Adversarial loss ($\mathbf{\mathcal{L}_{{ADV}}}$)}} In addition to metric loss, we train our network using adversarial loss. We train separate discriminator networks $\mathcal{D}_{{R}}$ and $\mathcal{D}_{{S}}$ for reverberant and clean speech respectively. Our $\mathcal{L}_{{ADV}}$ is described in Eq.~\ref{adv_loss}.
 {\small
 \belowdisplayskip 0.4\belowdisplayshortskip
\begin{equation}\label{adv_loss}
\begin{aligned}[b]
     & \operatorname{\mathcal{L}_{ADV} = \mathbb{E}[\max{(0,1-\mathcal{D}_{R}(\hat{\mathcal{S}_R})} + \max{(0,1-\mathcal{D}_{S}(\hat{\mathcal{S}_C})}],} \\
\end{aligned}
\end{equation}
}

We train speech dereverbaration RVQ codebook using VQ loss $\mathcal{L}_{{VQ2}}$~\cite{VQ1}. Eq.~\ref{gen_loss} presents our total generator loss $\mathcal{L}_{{Gen}}$. In Eq.~\ref{gen_loss}, $\lambda_{1}$ and $\lambda_{2}$ are the weights.
 {\small
 \belowdisplayskip 0.4\belowdisplayshortskip
\begin{equation}\label{gen_loss}
\begin{aligned}[b]
    \operatorname{ \mathcal{L}_{GEN}(x) =  \mathcal{L}_{METRIC} + \lambda_{1} \mathcal{L}_{ADV} + \lambda_{2} ( \mathcal{L}_{VQ1} + \mathcal{L}_{VQ2}),}
\end{aligned}
\end{equation}
}
\vspace{0.5mm}
 
{\noindent \textbf{Discriminator ($\mathcal{D_R},\mathcal{D_S}$).}} We use the multi-period discriminator network (MPD) proposed by HiFi-GAN~\cite{hi-fi_vocoder} and the multi-scale discriminator network (MSD) proposed in MelGAN~\cite{melGAN}. MPD effectively captures the periodic details by having several sub-discriminators, each handling different parts of the input audio. MSD captures consecutive patterns and long-term dependencies.

\subsection{Contrastive RIR-Image Pretraining (CRIP)}
\label{subsec:CRIP}
We propose CRIP, a model built on the fundamentals of CLIP~\cite{clip}, that learns a joint embedding space between the panoramic image ($\mathcal{I}_{P}$) and their corresponding RIRs. Similar to CLIP, CRIP employs two encoders, a pre-trained HorizonNet encoder~\cite{horizonet} $\mathcal{E}_{H}$ which serves as our $\mathcal{I}_{P}$ encoder, and the discriminator network proposed in FAST-RIR~\cite{fast-rir}, which serves as our RIR encoder $\mathcal{E}_{F}$. Formally, $\mathcal{E}_{H}$ takes as input $\mathcal{I}_P$ and outputs an embedding $\mathcal{I}_{EM} \in \mathbb{R}^{N \times 1024}$, and $\mathcal{E}_{F}$ takes as input $\mathcal{RIR} \in \mathbb{R}^{N \times 1 \times 4096} $ and output an embedding $\mathcal{R}_{EM} \in \mathbb{R}^{N \times 1024}$. Finally, we measure similarity by calculating the dot product between the two as follows:

{\small
\belowdisplayskip 0.8\belowdisplayshortskip
\begin{equation}\label{similarity} 
\begin{aligned}[b]
   & C_{r\mhyphen 2\mhyphen i} = \tau *\left(\mathcal{RIR} \cdot \mathcal{I}_{EM}^{\top}\right)  \\
    & C_{i\mhyphen 2\mhyphen r} = \tau *\left(\mathcal{I}_{EM} \cdot \mathcal{RIR}^{\top}\right)
\end{aligned}
\end{equation}
}

where $\tau$ is the temperature. This is followed by calculating the RIR-to-Image loss $\ell_{r\mhyphen 2\mhyphen i}$ and the  Image-to-RIR loss $\ell_{r\mhyphen 2\mhyphen i}$ as follows:
{\small
\belowdisplayskip 0.8\belowdisplayshortskip
\begin{equation}\label{similarity} 
\begin{aligned}[b]
\ell_{r\mhyphen 2\mhyphen i} = \frac{1}{N} \sum_{i=0}^N \log \operatorname{diag}(\operatorname{softmax}(C_{r\mhyphen 2\mhyphen i}))\\
\ell_{r\mhyphen 2\mhyphen i} = \frac{1}{N} \sum_{i=0}^N \log \operatorname{diag}(\operatorname{softmax}(C_{i\mhyphen 2\mhyphen r}))
\end{aligned}
\end{equation}
}
Finally, we optimize the average of both losses:
{\small
\belowdisplayskip 0.8\belowdisplayshortskip
\begin{equation}\label{similarity} 
\begin{aligned}[b]
\mathcal{L}=0.5 *\left(\ell_{r\mhyphen 2\mhyphen i}+\ell_{i\mhyphen 2\mhyphen r}\right)
\end{aligned}
\end{equation}
}

\noindent{\textbf{Why CRIP?}} Neural-network-based RIR estimators are known to inaccurately approximate the late components of the RIR as a sum of decaying filtered noise~\cite{fins}. Similarly, while our codec-based AV-RIR can accurately estimate the early components with structured impulsive patterns, it cannot precisely estimate late reverberation, which generally contains noise-like components. Thus, we propose CRIP to fill this gap. CRIP uses HorizonNet that captures room geometry/layout information~\cite{horizonet}. The late components of RIR depend on the geometry of the room~\cite{perez2019machine}. 
\vspace{0.4mm}


\noindent{\textbf{CRIP for AV-RIR Inference.}} During inference, for $\mathcal{I}_{P}$ of the target scene, we retrieve an RIR from a datastore $\mathcal{DS}$. The retrieval is performed by calculating cosine similarity between the CRIP embeddings of $\mathcal{I}_{P}$, denoted as $\mathcal{I}^{t}_{EM}$, and the RIR embeddings for all RIRs in datastore $\mathcal{DS}$. $\mathcal{DS}$ is generally a large collection of RIRs in the wild, which in our case is composed of synthetic RIRs, more details on which can be found in Section \ref{sec:experiments}. The final estimated RIR is obtained by replacing the late components of the original estimated RIR $\hat{RIR}$ with the late components of the retrieved RIR $\mathcal{RIR}_{CRIP}$. We perform hyper-parameter tuning to find the optimal number of samples $S$ from $\hat{RIR}$ to replace with $\mathcal{R}_{CRIP}$, and we found $S$ = 2000 to give us the best improvements. This whole process can be formalized as:
\vspace{-0.5em}
{\small
\belowdisplayskip 0.8\belowdisplayshortskip
\begin{equation}\label{similarity} 
\begin{aligned}[b]
&\mathcal{\hat{RIR}}[2000:4000] = \mathcal{RIR}_{CRIP}[2000:4000].
\end{aligned}
\end{equation}
}





%% file: sec/4_experiment.tex
\section{Experiments and Results}
\label{sec:experiments}


{\noindent\textbf{Datasets.}} For training and evaluation, we use the widely adopted SoundSpaces dataset~\cite{soundspaces1,soundspaces2}. The SoundSpaces dataset provides paired reverberant speech and its RIR. The data is sourced by convolving simulated RIRs with clean speech from the LibriSpeech~\cite{LibriSpeech} dataset. The RIRs are simulated using a geometrical acoustic simulation techniques~\cite{gsound,dynamic3} with environments taken from the Matterport3D dataset~\cite{Matterport3D}. To additionally evaluate how AV-RIR fairs in speech dereverberation in real-world scenarios, we use web videos in the filtered AVSpeech dataset~\cite{AV_Speech} proposed in VAM~\cite{visual_acoustic1}. Since AVSpeech does not have ground truth (GT) RIR, we only used it to evaluate our speech dereverberation pipeline. Our datastore $\mathcal{DS}$ comprises synthetic RIRs generated from SoundSpaces, excluding test set RIRs.

\looseness=-1
{\noindent \textbf{Hyperparameters.}} We train AV-RIR with a batch size of 16 for 400 epochs with only metric loss (Eq.~\ref{metric_loss}) and VQ loss ($\mathcal{L}_{{VQ1}}$, $\mathcal{L}_{{VQ2}}$). Later, we train with total loss (Eq.~\ref{gen_loss}) for 1K epochs. We use Adam Optimizer~\cite{kingma2014adam} with $\beta_1$ = 0.5, $\beta_2$ = 0.9 and learning rate $5\times10^{-5}$. For every 200K steps, we decay the learning rate by 0.5. 
\vspace{0.5mm}

\subsection{RIR Estimation}
\label{subsec:rir_estimation}
\looseness=-1
{\noindent\textbf{Evaluation Metrics.}} We quantitatively measure the accuracy of estimated RIR using standard room acoustic metrics. Reverberation time ($T_{60}$), direct-to-reverberant ratio (DRR), and early decay time (EDT) are the commonly used room acoustic statistics. We calculate the mean absolute difference between the acoustic statistics of estimated and ground truth (GT) RIRs as the error. $T_{60}$ measures the time taken for the sound pressure to decay by 60 decibels (dB), and EDT is 6 times the time taken for the sound pressure to decay by 10 dB. $T_{60}$ depends on the room size and room materials, and EDT depends on the type and location of the sound source~\cite{irgan}. DRR is the ratio between the sound pressure level of the direct sound source and the reflected sound~\cite{drr_book}. We also report the mean square difference (MSE) between the GT and estimated early component (EMSE) and late component (LMSE) of the RIR in time domain. We show the benefit of RIR estimated from our approach in SLP tasks in our supplementary.
\vspace{0.5mm}

{\noindent \textbf{Baselines.}} We compare the performance of AV-RIR with six other baselines. (1) \textbf{Image2Reverb~\cite{image2reverb}:} Predicts RIR from the camera-view image. (2) \textbf{Visual Acoustic Matching (VAM)~\cite{visual_acoustic1}:} Takes as input the source audio and the target environment image and outputs resynthesized audio matching the target environment. VAM does not explicitly estimate the RIR of the target environment. Therefore, we compare VAM only on our perceptual evaluation. (3) \textbf{FAST-RIR++ ~\cite{fast-rir}}: Takes as input the room geometry, source and listener positions, and $T_{60}$ to generate the RIR. We modified the architecture to the input panoramic image ($\mathcal{I}_{P}$) of the environment and estimate the RIR (FAST-RIR++). (4) \textbf{CRIP-only}~\textit{(ours)}: CRIP retrieves the closest RIR from the large synthetic dataset $\mathcal{DS}$ using a panoramic image $\mathcal{I}_{P}$ as input. We use it as a baseline to evaluate how much improvement our audio-visual network contributes. (5) \textbf{Filtered Noise Shaping Network (FiNS)~\cite{fins}:} An audio-only time domain RIR estimator from reverberant speech. (6) \textbf{S2IR-GAN ~\cite{s2ir}: } An audio-only GAN-based reverberant speech-to-IR estimator. 
\vspace{0.5mm}

{\noindent \textbf{Results.}} Table~\ref{table1} compares our approach AV-RIR with audio-only and visual-only baselines. We can see that the audio-only baselines outperform the visual-only baselines. Audio and visual cues provide complementary information for RIR estimation, and we can see a significant boost in performance in our AV-RIR, which inputs audio and visual cues. AV-RIR outperforms the SOTA audio-only approach S2IR-GAN by 36\%, 42\%, 63\%, 89\% and 98\% on $T_{60}$, DRR, EDT, EMSE, and LMSE, respectively.
\vspace{0.5mm}

\begin{table}[t]
    \setlength{\tabcolsep}{1.2pt}
	\renewcommand{\arraystretch}{0.75} 
	\caption{We compare the RIR estimated using our AV-RIR with prior visual-only method (Image2Reverb~\cite{image2reverb}) and audio-only methods (FiNS~\cite{fins} and S2IR-GAN~\cite{s2ir}). We perform an ablation study to show the benefit of each component of our network. } 
	\label{table1}
	\centering
       \resizebox{\columnwidth}{!}{
	\begin{tabular}{@{}llcccccc@{}}	
		\toprule
         & \textbf{Method}  & \textbf{$\mathbf{T_{60}}$ Error} & \textbf{DRR Error} & \textbf{EDT Error} & \textbf{EMSE}& \textbf{LMSE}   \\
          &  & \textbf{(ms)} & \textbf{(dB)} & \textbf{(ms)} & \textbf{(x$\mathbf{10^{-5}}$)}& \textbf{(x$\mathbf{10^{-5}}$)}   \\
		\midrule    &  Image2Reverb~\cite{image2reverb}      & 131.7& 4.94        &     382.1     & 4907 & 1126 \\
                        &  FAST-RIR++~\cite{fast-rir}      & 126.4 &  3.62     &  334.2     & 2630 & 990 \\
                        &  FiNS~\cite{fins}             & 87.7        & 3.30           &   235.7        & 924  & 561   \\
                        &  S2IR-GAN~\cite{s2ir}             & 63.1  & 3.04         &   168.3     & 730 & 310 \\
            \midrule
        & AV-RIR (Audio-Only) &   88.8     & 2.96 & 122.4 &   176&   51  \\
        & AV-RIR w Random &   77.6     & 2.67 & 109.2 &   124 &   6 \\
        & AV-RIR w/o CRIP &   61.7     & 2.07  & 79.8  &   79 &   42 \\
        & AV-RIR w/o Geo-Mat &  55.7       & 1.98  & 74.1 &   104 &   6 \\
        & AV-RIR w/o Multi-Task &  77.8        & 2.56 & 105.4 &  144 &  6  \\
      \multirow{-6}{*}{\begin{turn}{90}\textbf{ABLATION}\end{turn} }   & AV-RIR w/o STFT Loss &  59.4       & 1.94  & 77.2&   123  &   6 \\
        \midrule    
        & CRIP-only \textit{(ours)}& 118.9          & 3.14 & 298.4  &212 &6  \\
        & \myccone  \textbf{AV-RIR} \textit{ \textbf{(ours)}}& \myccone \textbf{40.2  }        & \myccone\textbf{1.76 } & \myccone\textbf{62.1 }& \myccone \textbf{82 }  & \myccone \textbf{6 }  \\
 
        \midrule
  
		
	\end{tabular}}
\end{table}

\begin{table}[t]
\caption{\small Performance comparison of AV-RIR with audio-only baselines (marked with $\ddagger$) and audio-visual baselines (marked with $\ast$) on the SLP tasks on the SoundSpaces dataset. ``Reverberant'' refers to clean speech convolved with ground-truth RIR. We also report RTE for real-world audio from the AVSpeech dataset.} 
\centering
\resizebox{\columnwidth}{!}{%
\begin{tabular}{@{}ll|ccccc|c@{}}
\toprule
\multicolumn{2}{l|}{}                                                & \textbf{Speech Recognition$^{\dagger}$}                                          & \textbf{Speaker Verification$^{\dagger}$}                                       & \textbf{RTE*  ↓}                          \\

\multicolumn{2}{l|}{\multirow{-2}{*}{\textbf{Method}}}                                  & \textbf{WER (\%) ↓}                                                & \textbf{EER (\%) ↓}                                                     & \textbf{(in sec)}                      \\ \midrule

\multicolumn{2}{l|}{Clean (Upper bound)}& 2.89        & 1.53                                          & -                                      \\ \midrule
\multicolumn{2}{l|}{Reverberant} & 8.20  & 4.51 & 0.382                                  \\
\multicolumn{2}{l|}{MetricGAN+~\cite{fu2021metricgan+}{$\ddagger$}} & 7.48 (+9\%)  & 4.67 (-4\%)  & 0.187 (+51\%)\\
\multicolumn{2}{l|}{DEMUCS~\cite{defossez20_interspeech}{$\ddagger$}} & 7.97 (+3\%)  & 3.82 (+15\%)  & 0.129 (+66\%)\\
\multicolumn{2}{l|}{HiFi-GAN~\cite{hi-fi}{$\ddagger$}} & 9.31 (-14\%)  & 4.32 (+4\%) & 0.196 (+49\%)                                  \\
\multicolumn{2}{l|}{WPE~\cite{nakatani2010speech}{$\ddagger$}}  & 8.43 (-3\%) & 5.90 (-31\%)   & 0.173 (+55\%)                                  \\
\multicolumn{2}{l|}{VoiceFixer~\cite{liu22y_interspeech}{$\ddagger$}} & 5.66 (+31\%)  & 3.76 (+16\%)  & 0.121 (+68\%)\\
\multicolumn{2}{l|}{SkipConvGAN~\cite{skip-convgan}{$\ddagger$}}& 7.22 (+12\%)  & 4.86 (-8\%)    & 0.119  (+69\%)                                \\
\multicolumn{2}{l|}{Kotha \textit{et al.}~\cite{kotha}{$\ddagger$}}& 5.32 (+35\%)  & 3.71 (+17\%)    & 0.124  (+68\%)                                 \\
\midrule
\multicolumn{2}{l|}{VIDA~\cite{AV_changan}{$\ast$}} & 4.44 (+46\%)   & 3.97 (+12\%)   & 0.155 (+59\%)                                 \\ 
\multicolumn{2}{l|}{AdVerb~\cite{AV_sanjoy}{$\ast$}}    & \textbf{3.54 (+57\%)}   & 3.11 (+31\%)   & 0.101 (+74\%)    \\
\midrule
                   & AV-RIR (Audio-Only)                                           &   5.24  (+36\%)                                                                &        2.67  (+41\%)                                                          &    0.055 (+86\%)                             \\
                   & AV-RIR w Random  Image                                                &   4.85  (+41\%)                                                               &        2.56    (+43\%)                                                   &  0.049  (+87\%)                              \\
                  
                   & AV-RIR w/o  Multi-Task                                       &   4.57 (+44\%)                                                                &         2.55  (+43\%)                                                      &      0.048  (+87\%)                          \\
                   & AV-RIR w/o  CRIP                                                        &   4.67   (+43\%)                                                             &         2.66  (+41\%)                                                      &  0.049  (+87\%)                              \\
                    & AV-RIR w/o  Geo-Mat                                                    &    4.54 (+45\%)                                                             &            2.21(+51\%)                                                     &   0.044 (+88\%)                              \\
 \multirow{-6}{*}{\begin{turn}{90}\textbf{ABLATION}\end{turn} }                  & AV-RIR w/o  MEL Loss                                                    &    4.44  (+46\%)                                                             &            2.44 (+46\%)                                                    &   0.048  (+87\%)                             \\
\midrule
\multicolumn{2}{l|}{ \myccone \textbf{AV-RIR} \textit{\textbf{(ours)}}}                                                            & \myccone 4.17 (+49\%)                                                                   & \myccone    \textbf{2.02 (+55\%) }                                                              & \myccone    \textbf{0.042 (+89\%)}                 
\\
\bottomrule
\end{tabular}%
}

\label{tab:table2}
\end{table}

\begin{figure*}[t] 
\centering
\includegraphics[width=0.9\textwidth]{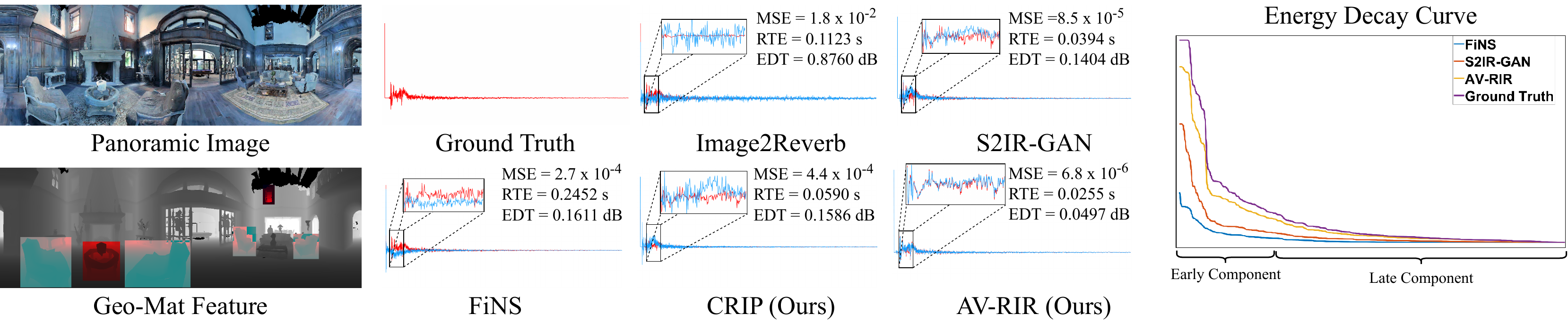}
\caption{\small \textbf{Qualitative Results}. (Left) We show the Geo-Mat feature generated using our approach. The cushion chairs with a similar material absorption property are represented in green. The table and window with similar material are represented in red. (Right) We plot the time-domain representation of the RIRs estimated using prior methods and our approach with the ground truth (GT) RIR (GT: Red, Estimated: Blue). We also report the MSE (Eq.~\ref{rir_loss}), $T_{60}$ error (RTE), and EDT error (EDT). It can be seen that the RIR estimated using our AV-RIR matches closely with GT RIR when compared with the baseline. Also, we can see that the RIR retrieved from our CRIP has similar late components as the GT RIR. However, the early component of the retrieved RIR (shown in zoom) significantly differs from the GT. Our full AV-RIR pipeline estimates the early components of the RIR using audio-visual features and adds the late component of the RIR from our CRIP to accurately predict the full RIR. The energy decay curve (EDC) depicts the energy remaining in the RIR over time~\cite{EDC1}. We can see that the EDC of the late component of RIR estimated from AV-RIR (yellow) matches closely with the GT RIR (purple). }
\label{fig:eval_fig}
\end{figure*}
\subsection{Speech Dereverberation}
\label{subsec:speech_enhancement}

{\noindent\textbf{Evaluation Metrics.}}
We evaluate speech dereverberation using our AV-RIR on two downstream tasks: Automatic Speech Recognition (ASR) and Speaker Verification (SV). We use standard metrics such as Word Error Rate (WER) to measure ASR performance and Equal Error Rate (EER) to measure the SV. Following prior works~\cite{AV_changan,AV_sanjoy}, we used pre-trained models from SpeechBrain~\cite{speechbrain} to evaluate the ASR and SV performance. AVSpeech dataset does not have parallel clean speech to perform ASR and SV tasks and we use the Reverberation Time Error (RTE) metric~\cite{AV_sanjoy} to evaluate the estimated clean speech from our AV-RIR.

{\noindent\textbf{Baselines.}} We compared the speech dereverberation using our approach with the following prior audio-only and audio-visual speech enhancement networks. (1) \textbf{Audio-Only: } WPE~\cite{nakatani2010speech} is a statistical-model-based speech dereverberation network that cancels late reverberation without the knowledge of RIR. MetricGan+~\cite{fu2021metricgan+}, DEMUCS~\cite{defossez20_interspeech}, HiFi-GAN~\cite{hi-fi}, VoiceFixer~\cite{liu22y_interspeech}, SkipConvGAN~\cite{skip-convgan} and Kotha \textit{et al.}~\cite{kotha} are learning-based speech dereverberation networks. (2) \textbf{Audio-Visual :} VIDA~\cite{AV_changan} is the first audio-visual speech dereverberation network that takes $\mathcal{I}_{P}$ as visual cues. Recently, geometry-aware AdVerb~\cite{AV_sanjoy} has been shown to achieve SOTA results in downstream speech tasks. 
\vspace{0.5mm}

{\noindent \textbf{Results.}}
Table~\ref{tab:table2} shows the benefits of speech dereverberation from our AV-RIR in the ASR and SV task. We can see that AV-RIR outperforms SOTA network AdVerb in SV tasks by 35\% and outperforms all the baselines except for AdVerb in the ASR. To evaluate the robustness of AV-RIR, we test our network on recorded speech not used for training in the AVSpeech using the RTE metric. We can see that our work outperforms all the baselines by 60\%. 
\vspace{0.5mm}

\looseness=-1
{\noindent \textbf{Ablation Study.}} We perform a comprehensive ablation study to show the benefit of different components of our AV-RIR. (1)~\textbf{Multi-task learning.} To prove the effectiveness of the multi-task learning approach, we train the branches separately (AV-RIR w/o Multi-Task). Table~\ref{table1} and Table~\ref{tab:table2} show that our multi-task learning approach benefits both tasks mutually. While the RIR estimation performance improves by 31\% - 48\%, speech dereverberation performance improves by 13\% - 21\%. (2)~\textbf{CRIP.} Table~\ref{table1} shows that adding CRIP during inference for late reverberation improves the late component MSE (LMSE) by 86\%. (3)~\textbf{Geo-Mat feature.} Table~\ref{table1} shows that the Geo-Mat feature improves RIR estimation accuracy by 11\%  - 28\%. (4)~\textbf{Visual cues.} To prove the effectiveness of visual cues, we discard them while training and directly pass RIR and speech encoder outputs to RVQ. Additionally, we also discard CRIP and directly estimate the full RIR. Table~\ref{table1} and Table~\ref{tab:table2} show that AV-RIR outperforms our audio-only AV-RIR variation in RIR estimation tasks by 41\% - 55\% and speech dereverbation task by around 24\%.

\subsection{Perceptual Evaluation}
In Table~\ref{table3} we report scores for perceptual evaluation of our estimated RIRs. For each environment, we provide the ground truth (GT) speech, generated speech using Image2Reverb~\cite{image2reverb}, VAM~\cite{visual_acoustic1}, our AV-RIR and the environment image. The participants were asked to select the generated speech that sounds closer to the GT speech.
\vspace{0.5mm}

Deatils on initial pre-screening of participants is described in the Appendix. We select 6 scenes with varying complexity with $T_{60}$ ranging from 0.2 seconds to 0.7 seconds. We can see that, irrespective of the $T_{60}$ and the environment complexity, 56\% to 79\% of the participants said that the generated speech from AV-RIR closely matched the GT speech.

\begin{table}[t]
	\caption{\textbf{Perceptual Evaluation}. Participants find that the reverberant speech generated using our AV-RIR is closer to GT reverberant speech when compared to visual-only baselines.} 
	\label{table3}
	\centering
       \resizebox{\columnwidth}{!}{%
	\begin{tabular}{cccccc}	
		\toprule
     & \textbf{Scene}  & \textbf{$\mathbf{T_{60}}$} & \textbf{Image2Reverb~\cite{image2reverb}} & \textbf{VAM\cite{visual_acoustic1}}   & \myccone \textbf{AV-RIR (Ours)} \\
        
		\midrule    
            &Scene 1        & 0.22 &  2\%       &    19\%      &\myccone  \textbf{79\%}  \\
            &Scene 2        & 0.31 &  16\%       &   26\%       &\myccone \textbf{58\%}  \\
            &Scene 3        & 0.35 &  14\%       &   16\%     &\myccone  \textbf{70\%}   \\
            &Scene 4        & 0.38 & 5\%        &    40\%       &\myccone  \textbf{56\%} \\
            &Scene 5        & 0.47 & 16\%        &   16\%       &\myccone  \textbf{67\%} \\
            &Scene 6        & 0.65 & 14\%        &  12\%       &\myccone  \textbf{74\%}   \\
\bottomrule
	\end{tabular}}
	\vspace{-0.5mm}
\end{table}

%% file: sec/5_conclusion.tex
\section{Conclusion, Limitations and Future Work}
\label{sec:conclusions}

We propose AV-RIR, a novel multi-modal multi-task learning approach for RIR estimation. AV-RIR leverages both audio and visual cues using a novel neural codec-based multi-modal architecture and solves speech dereverberation as the auxiliary task. We also propose Contrastive RIR-Image Pre-training (CRIP), which improves late reverberation components in estimated RIR using retrieval. Both quantitative metrics and perceptual studies show that our AV-RIR significantly outperforms all the baselines. We evaluate the speech dereverberation performance on the recorded AVSpeech dataset not used for training and observe that our approach outperforms the baselines by 60\%. 

AV-RIR assumes stationary single-talker input speech or single-source audio without noise. Future work aims to tackle multi-channel RIR estimation and RIR estimation from noisy, multi-source environment with moving sources. 


%% file: sec/6_supplementary.tex
\section{Table of Contents:}
\label{sec:supplementary}

In the Supplementary Material, we provide additional information about:

\begin{itemize}
  \item The qualitative results of our AV-RIR via a supplementary video\footnotemark{}.
  \item Quantitative comparison of AV-RIR on far-field Automatic Speech Recognition with other baselines.
  \item Additional details on our datasets and baselines used for evaluation.
  \item Additional details of our user study.
  \item Information about Societal Impact of AV-RIR.
\end{itemize}

\footnotetext{\url{https://www.youtube.com/watch?v=tTsKhviukAE}}

\subsection{Supplementary Video}

We provide a supplementary video showing the qualitative results of RIR estimation with AV-RIR when applied to three different tasks. In addition, we compare the enhanced speech from our AV-RIR with ground truth clean speech. We also demonstrate our approach's failure cases in the supplementary video.

The RIRs estimated from our approach are evaluated in three practical tasks, that are:
\vspace{0.5mm}

\noindent \textbf{Novel View Acoustic Synthesis :} In the novel view acoustic synthesis task, given the audio-visual input from the source viewpoint, we modify the reverberant speech from the source viewpoint to sound as if it is recorded from the target viewpoint. We use reverberant speech as audio input, and the panoramic image and our proposed Geo-Mat feature as our visual input. We use SoundSpaces~\cite{soundspaces1} dataset to perform this task.

To perform this task, we estimate the enhanced speech using audio-visual input from the source viewpoint. We estimate the RIR corresponding to the target viewpoint from audio-visual input. We convolve the enhanced speech from the source viewpoint with RIR from the target viewpoint to make the speech from the source viewpoint sound as if it is recorded from the target viewpoint.
\vspace{0.5mm}

\noindent \textbf{Visual-Acoustic Matching :} In the visual-acoustic matching task, we resynthesize the speech from the source environment to match the target environment. We combine the enhanced source environment speech from AV-RIR and the estimated RIR from the target environment to perform this task. Convolving the estimated RIR with clean speech leads to synthesizing speech from the source environment to match the target environment. All our experiments on this task are performed on the SoundSpaces~\cite{soundspaces1} dataset.
\vspace{0.5mm}


\noindent \textbf{Voice Dubbing :} Voice dubbing is replacing dialogue in one language with another in a video. Voice dubbing is commonly used to dub movies from one language to another. To test the robustness of RIR estimation from our AV-RIR, we estimated RIR using our AV-RIR on recorded video clips on YouTube. We chose two English video clips in the AVSpeech dataset~\cite{AV_Speech}. We dubbed the video clips with French clean speech from Audiocite~\cite{audiocite}. We convolved the French clean speech with the estimated RIR from the YouTube clip to match the room acoustics of French dialogue with the original English dialogue. We replaced the English dialogue with modified French dialogue using our approach.
\vspace{0.5mm}

\noindent \subsection{Far-field Automatic Speech Recognition}
In order to evaluate the effectiveness of RIR estimated from our AV-RIR, we performed a Kaldi Far-field Automatic Speech Recognition (ASR) experiment using a modified KALDI ASR recipe\footnote{\url{https://github.com/RoyJames/kaldi-reverb/tree/ami/}}. For our experiment, we use the AMI corpus~\cite{ami}. The AMI corpus has 100 hours of meeting recording. The meeting is recorded using both an individual headset microphone (IHM) and a single distant microphone (SDM). The IHM data has a high signal-to-distortion ratio when compared to the SDM data. Therefore, IHM data can be considered as clean speech.

To evaluate the benefit of RIRs estimated from our AV-RIR, we take a subset of SDM data with 300 speech samples and estimate the RIRs of the subset of SDM data. We create synthetic reverberant speech data by convolving clean speech from IHM data with the estimated RIR. We train the KALDI ASR recipe with and without modifying the IHM using our audio-only AV-RIR. We evaluated the audio-only version of our AV-RIR because there are no corresponding visual inputs in the AMI corpus. We test the trained model on far-field SDM data. We use word error rate as our metric to evaluate the performance of the speech recognition system. A lower word error rate indicates improved performance.

Modifying IHM data using our audio-only AV-RIR will bridge the domain gap between the training and test data. From Table~\ref{table_ami} we can see that modifying the IHM data with our audio-only AV-RIR improves the word error rate by 12\%.

\begin{table}[t]
  \setlength{\tabcolsep}{1.8pt}
	\renewcommand{\arraystretch}{0.7} 
	\caption{Far-field ASR results. We train the Kaldi ASR recipe with and without modified IHM data and test on SDM data. We modify the IHM data by convolving RIR estimated using our audio-only AV-RIR.}
	\label{table_ami}
	\centering
	\begin{tabular}{@{}llc@{}}	
		\toprule
		\multicolumn{2}{l}{{\textbf{Training Dataset}}}
			& \textbf{Word Error Rate $\downarrow$} [\%]\\
		\midrule
		&IHM without Modification  & 64.2 \\
		&\textbf{IHM \boldmath$\circledast$ AV-RIR (ours)} & \textbf{52.1} \\
			
		\bottomrule
	\end{tabular}
	\vspace{-0.3cm}
\end{table}

\subsection{Additional details on our datasets and baselines}

\subsubsection{SoundSpaces dataset:} We trained and test our network on SoundSpaces dataset~\cite{soundspaces1}. The SoundSpaces dataset comes with a non-overlapping clean speech from the LibriSpeech dataset~\cite{LibriSpeech} and synthetic reverberant speech. The synthetic reverberant speech is simulated using the geometric acoustic simulator in the SoundSpaces platform~\cite{soundspaces1,soundspaces2} for 82 Matterport~\cite{Matterport3D} 3D environments. SoundSpaces can simulate highly realistic RIR for any arbitrary camera views and microphone positions by considering direct sounds, early reflections, late reverberations, material and air absorption properties, etc. The panoramic images in the SoundSpaces dataset contain 3D humanoids of the same gender as the speaker in each data. In some data, the speaker is out of view and will not be visible in the panoramic image. The sound spaces dataset has 49,430/2700/2,600 train/validation/test samples respectively.

\subsubsection{AVSpeech Web Video dataset:}
To evaluate the robustness of our approach, we evaluate our RIR estimation and Speech enhancement approach using a subset of 1000 speeches in AVSpeech dataset~\cite{AV_Speech} proposed in Visual Acoustic Matching paper~\cite{visual_acoustic1}. The filtered dataset contains 3-10 seconds YouTube clips with reverberant audio recordings. Also, the filtered dataset microphone and the camera are co-located and placed at a different position than the sound sources. The cameras in the filtered videos are static.

\subsubsection{Speech enhancement baselines:}
\textbf{MetricGAN++\cite{fu2021metricgan+}:} MetricGAN++ is an improvised version of the MetricGAN framework where the discriminator network is trained with noisy speech. We use the implementation of MetricGAN in Speechbrain for our comparison~\cite{speechbrain}.
\vspace{0.5mm}

\noindent \textbf{DEMUCS~\cite{defossez20_interspeech}:} DEMUCS is the music source separation architecture in the time-domain modified into a time-domain speech enhancer. DEMUCS can work in real-time on consumer-level CPUs.
\vspace{0.5mm}

\noindent \textbf{HiFi-GAN~\cite{hi-fi}:} HiFi-GAN is GAN-based architecture trained on multi-scale adversarial loss in both the time domain and time-frequency domain to enhance real-world speech recording to studio quality.
\vspace{0.5mm}

\noindent \textbf{WPE~\cite{nakatani2010speech}:} WPE is a statistical model-based speech dereverberation approach. WPE can perform dereverberation by removing late reverberation in a reverberant speech signal.
\vspace{0.5mm}

\noindent \textbf{VoiceFixer~\cite{liu22y_interspeech}:} Voice-fixer is a two-stage speech dereverberation approach. The analysis stage of the VoiceFixer is modelled using ResUNet and the synthesis stage is modelled using TF-GAN.
\vspace{0.5mm}

\noindent \textbf{SkipConvGAN~\cite{skip-convgan}:} SkipConGAN is the GAN-based speech enhancement architecture where the Generator network estimates the complex time-frequency mask and the discriminator network helps to restore the formant structure in the synthesized enhanced speech.
\vspace{0.5mm}

\noindent \textbf{Kotha \textit{et al.}~\cite{kotha}:} This speech enhancement network integrates the complex-valued TFA module with the deep complex convolutional recurrent network to improve the overall speech quality of the enhanced speech.
\vspace{0.5mm}

\noindent \textbf{VIDA~\cite{AV_changan}:} VIDA is the audio-visual speech dereverberation network that enhances reverberant speech. Visual input gives valuable information about the room geometry, materials and speaker positions.
\vspace{0.5mm}

\noindent \textbf{AdVerb~\cite{AV_sanjoy}:} Adverb is a geometry-aware cross-modal transformer architecture, that predicts the complex ideal ratio mask. Clean speech is estimated by applying the complex ideal ratio mask to reverberant speech.
\vspace{0.5mm}
\begin{figure*}[t] 
	\centering
	\includegraphics[width=\linewidth]{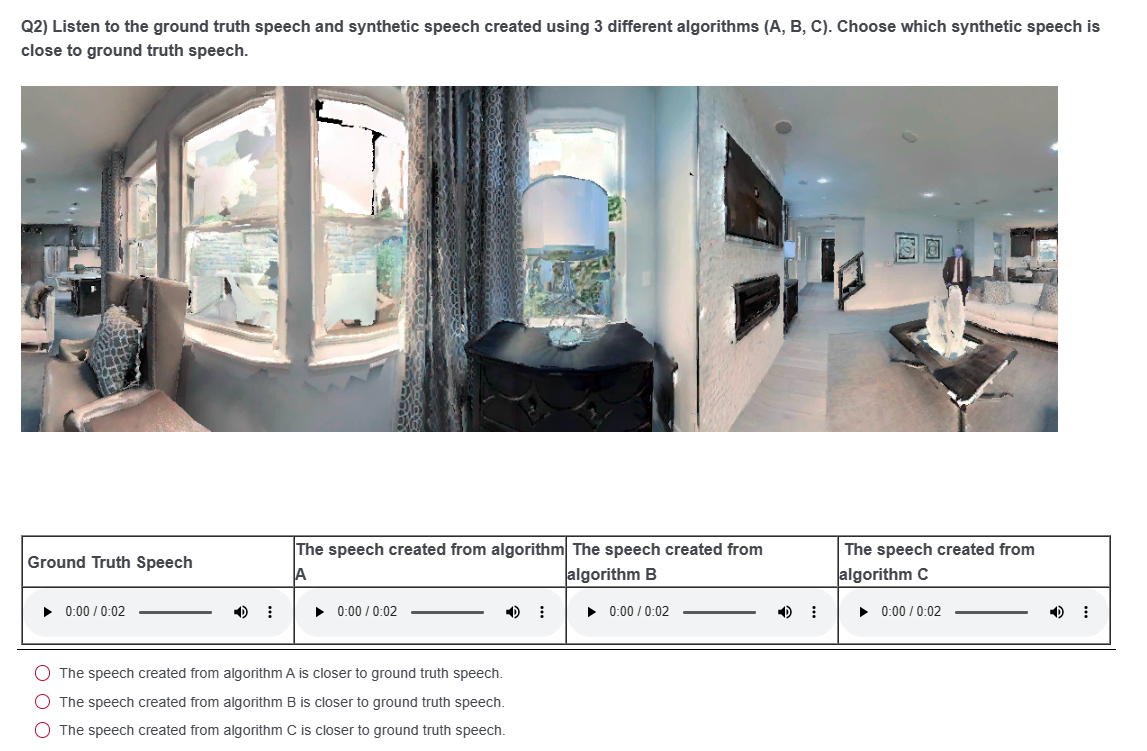}
  
	\caption{User study interface. We created 3 synthetic speech using our AV-RIR, Image2Reverb~\cite{image2reverb} and Visual Acoustic Matching~\cite{visual_acoustic1} and asked the participants, which synthetic reverberant speech matches closely with ground-truth reverberant speech. For each question, we randomly shuffle the order of the synthetic reverberant speech from different approaches.} 
	\label{userstudy}
\end{figure*}
\subsection{Additional details on User Study}

We performed our user study on 50 participants. We only allow participants to perform the survey on a laptop or desktop with headphones to get accurate results. Among the 50 responses, we filtered out noisy responses from our first questions. In the first question, we ask the participants which of the three synthetic reverberant speech matches closely to the ground-truth speech. We place ground truth speech among the synthetic speech and expect the participants with good hearing to choose the ground truth speech. We only counted the responses of 43 participants who chose the ground truth speech.

Out of 50 participants, 33 are male and 17 are female. The six participants are aged between 18-24 years, 28 participants are between 25-34 years and 16 participants are older than 34 years. Figure~\ref{userstudy} shows the second question from our user study interface. \\


\subsection{Societal Impact}
Our model to estimate the RIR and enhance speech can have positive impacts on real-world applications. For example, the model can give an immersive experience in AR/VR applications and improve voice dubbing in movies. Also, our can be useful for different speech processing applications such as automatic speech recognition systems, telecommunication systems etc. We trained and evaluated our network on open-sourced publicly available datasets.

We got the certification and license to perform user studies from the Institutional Review Board and we followed their protocols. We did not collect any personal information from the participants. 